\documentclass[aps,prd,twocolumn,superscriptaddress]{revtex4-1}

\usepackage{ulem}
\usepackage{amssymb}
\usepackage{amsmath}
\usepackage{dcolumn}
\usepackage{graphicx}
\usepackage{mathrsfs}
\usepackage{subfigure}
\usepackage{booktabs}
\usepackage{color}
\usepackage{docmute}
\usepackage{hyphenat}
\usepackage{CJKutf8}
\usepackage{multirow}
\usepackage{array}
\usepackage{threeparttable}
\usepackage[mathscr]{euscript}

\usepackage{graphicx}
\usepackage{subfigure}
\usepackage{tikz}
\usetikzlibrary{positioning}
\usepackage{xcolor}

\newcommand{\bea}{\begin{eqnarray}}
\newcommand{\eea}{\end{eqnarray}}
\newcommand{\beq}{\begin{equation}}
\newcommand{\eeq}{\end{equation}}
\newcommand{\nn}{\nonumber}
\def\/{\over}

\usepackage{url}
\usepackage[colorlinks]{hyperref}
\hypersetup{
	plainpages=true,
	breaklinks=true,
	hypertexnames=false,
	pageanchor=true,
	bookmarksnumbered=true,
	colorlinks=true,
	linkcolor={blue},
	citecolor={red},
	urlcolor={blue},
	anchorcolor={black}
}
\usepackage{orcidlink}

\allowdisplaybreaks[4]
\begin{document}
\title{Significant modifications of Lamb shift at small centripetal accelerations}
\author{Yan Peng\orcidlink{0009-0004-7100-1307}}
\affiliation{School of Fundamental Physics and Mathematical Sciences, Hangzhou Institute for Advanced Study, University of Chinese Academy of Sciences, No. 1 Xiangshan Branch, Hangzhou 310024, China}
\affiliation{Department of Physics, Key Laboratory of Low Dimensional Quantum Structures and Quantum Control of Ministry of Education, and Hunan Research Center of the Basic Discipline for Quantum Effects and Quantum Technologies, Hunan Normal University, Changsha, Hunan 410081, China}

\author{Jiawei Hu\orcidlink{0000-0002-7668-7537}}
\email[Corresponding author: ]{jwhu@hunnu.edu.cn}
\affiliation{Department of Physics, Key Laboratory of Low Dimensional Quantum Structures and Quantum Control of Ministry of Education, and Hunan Research Center of the Basic Discipline for Quantum Effects and Quantum Technologies, Hunan Normal University, Changsha, Hunan 410081, China}

\author{Hongwei Yu\orcidlink{0000-0002-3303-9724}}
\email[Corresponding author: ]{hwyu@hunnu.edu.cn}
\affiliation{Department of Physics, Key Laboratory of Low Dimensional Quantum Structures and Quantum Control of Ministry of Education, and Hunan Research Center of the Basic Discipline for Quantum Effects and Quantum Technologies, Hunan Normal University, Changsha, Hunan 410081, China}

\begin{abstract}
We investigate the Lamb shift of centripetally accelerated atoms coupled to electromagnetic vacuum fluctuations. Focusing on a very small orbital radius (so that the tangential speed remains nonrelativistic and the proper centripetal acceleration can be extremely small),
we show that the resulting level shift is intrinsically anisotropic and depends sensitively on the atomic polarization direction.  For atoms polarizable along the rotation axis, the leading noninertial contribution enters only at second order in the orbital radius and can slightly increase the energy-level spacing. For atoms polarizable perpendicular to the rotation axis, the noninertial contribution appears already at zeroth order in the radius and always increases the energy-level spacing. Remarkably, when the angular velocity greatly exceeds the transition frequency, the rotation-induced correction can become comparable in magnitude to the inertial Lamb shift, indicating that circular motion can significantly modify the Lamb shift even in the regime of very small centripetal accelerations.

\end{abstract}

\maketitle

\section{Introduction} \label{sec1}

According to the Heisenberg uncertainty principle~\cite{Heisenberg27}, quantum fields exhibit intrinsic fluctuations even in the vacuum state. Crucially, the perception of these fluctuations is not absolute but depends on the observer’s state of motion.
A paradigmatic example is the Fulling-Davies-Unruh effect (hereafter referred to as the Unruh effect), which states that the vacuum seen by an inertial observer appears to a uniformly accelerated observer as a thermal bath  with temperature proportional to the observer’s proper acceleration~\cite{Davies75,Unruh76,Fulling73}.
As a result, a quantum system  undergoing uniform acceleration may undergo spontaneous excitation even when it is initially in its ground state---a direct manifestation of the Unruh effect. This constitutes the basic operating principle of the Unruh-DeWitt particle detector model~~\cite{Unruh76,DeWitt1980}. Owing to the fundamental significance of the Unruh effect and its deep connection to the Hawking radiation~\cite{Hawking74,Hawking75} through the equivalence principle, the response of a uniformly accelerated Unruh-DeWitt detector has been extensively investigated; see, e.g., Refs.~\cite{Audretsch94,Takagi86,Zhu06}. In particular, for atoms undergoing uniform linear acceleration, the spontaneous excitation rate  is exponentially suppressed relative to the emission rate by a factor $e^{-2\pi\omega_0/a}$, where $\omega_0$ and $a$ are the atomic transition frequency and proper acceleration, respectively. Hence, for a single uniformly accelerated detector in free space at small accelerations, the excitation probability is extremely small, rendering direct detection experimentally challenging.
Nevertheless, recent studies show that this limitation can be partially alleviated. The efforts include, for example, the enhancement of acceleration-induced transitions and Lamb shifts via cavity resonance~\cite{Stargen2022,Arya2024}, and the use of cavity-assisted superradiance of a collection of atoms~\cite{Deswal2025}.

Uniform linear acceleration is, however, only one representative form of noninertial motion.
Another fundamental and experimentally accessible case is centripetal (circular) acceleration.
In the vacuum defined by inertial observers, a detector undergoing centripetal acceleration also registers excitations, even though the associated spectrum is generally nonthermal~\cite{Letaw1980,Bell83,Hacyan1986,Bell87,Kim1987,Unruh98}.  This ``circular Unruh effect'' likewise originates from acceleration-modified vacuum fluctuations and has motivated sustained interest in the radiative properties of atoms (and Unruh-DeWitt detectors) on circular orbits, coupled to scalar or electromagnetic vacuum fields~\cite{Letaw1981,Korsbakken2004,Good2020,Takagi1984,DeLorenci99,yu24,Peng25,Takagi86,Biermann2020,Bunney23,Rogers88,Davies1996,Lorenci2000,Bunney23-2}.
In particular, recent work shows that, when the rotational angular velocity exceeds the atomic transition frequency, the excitation rate  can become comparable to the emission rate, even when the centripetal acceleration is extremely small, enabled by a sufficiently  small orbital radius~\cite{yu24}. This suggests that the circular version of the Unruh effect can be significant even in the regime of very small centripetal accelerations and raises the broader possibility that circular motion may provide an experimentally favorable route for probing acceleration-induced vacuum effects in regimes where linear-acceleration signatures are essentially unobservable.

Among observables that encode atom-vacuum interactions, the Lamb shift holds particular significance~\cite{Lamb47}. As a cornerstone of quantum electrodynamics,  it is measured with extraordinary precision in modern spectroscopy~\cite{Bezginov19}. This precision has made the Lamb shift not only a stringent test of quantum electrodynamics, but also  a sensitive  probe of subtle modifications to vacuum fluctuations induced by noninertial motion and spacetime curvature \cite{Audretsch95,Audretsch95b,Passante98,Arya23,Zhou10,Zhou12}.
Given the nontrivial  transition  behavior found for centripetally accelerated atoms~\cite{yu24},
it is therefore natural to ask how the Lamb shift is altered under circular motion and whether it exhibits similarly distinctive dependence on the kinematic parameters.

In this work, we present a systematic investigation of the Lamb shift of a centripetally accelerated atom interacting with the vacuum electromagnetic field in free space.  We focus on the nonrelativistic regime with a very small orbital radius, such that the linear speed remains much smaller than the speed of light, and we allow the centripetal acceleration to be extremely small.
We show that circular motion nevertheless imprints characteristic, generally anisotropic corrections to the level shift, with contributions that depend on polarization directions relative to the orbital plane.
The paper is organized as follows. In Sec.~\ref{sec2}, we briefly review the general formalism for calculating the Lamb shift within the open quantum system framework. In Sec.~\ref{sec3}, we present the Lamb shift for an inertial atom in free space as a reference. In Sec.~\ref{sec4}, we derive the Lamb shift for a centripetally accelerated atom and analyze in detail the anisotropic contributions associated with specific polarization directions. Finally, we summarize in Sec.~\ref{sec5}.

\section{The basic formalism} \label{sec2}
For a two-level atom coupled to fluctuating electromagnetic fields in vacuum, the total Hamiltonian of the combined atom-field system can be expressed as $ H=H_A+H_F+H_I$. In the laboratory frame, the atomic Hamiltonian takes the form
$H_A = \frac{1}{2}\hbar \omega \sigma_3$, where $\hbar$ is the reduced Planck constant and $\sigma_3$ denotes the Pauli matrix. Here, $\omega$ is the transition frequency of the two-level atom observed in the laboratory frame, which characterizes the internal dynamics of the atom with respect to the laboratory coordinate time $t$. This frequency is related to the proper transition frequency $\omega_0$ via $\omega = \omega_0/\gamma$, where $\gamma = (1 - v^2/c^2)^{-1/2}$ is the Lorentz factor, $v$ is the linear velocity of the atom, and $c$ is the speed of light. Notably, the energy level spacing $\hbar \omega_0$ is an intrinsic property of the atom defined in its proper frame, where the internal state evolves with respect to the proper time $\tau$. Since $d\tau = dt/\gamma$ due to relativistic time dilation, expressing the same internal evolution in terms of the laboratory time coordinate $t$ leads to $\omega = \omega_0/\gamma$. 
The term $H_F$ represents the Hamiltonian of the free electromagnetic field, the specific detail of which is not required here. The interaction Hamiltonian in the laboratory frame can be expressed as $H_{I}= - \frac{1}{\gamma}{d}^{\mu} F_{\mu\nu} u^{\nu}$, where ${d}^{\mu}$ is the atomic four-vector electric dipole moment operator, which in the atom’s proper frame is given by $ (0, {d}_{\rho}, {d}_{\phi}, {d}_z) $, $F_{\mu\nu}$ denotes the electromagnetic field tensor, and $u^{\nu}$ denotes the atomic four-velocity. For further details regarding the interaction Hamiltonian of a centripetally accelerated atom in the laboratory frame, see Appendix~\ref{sec:S1}.

Initially, the total system is assumed to be prepared in the state
$\rho_{\rm{ tot}} = \rho(0) \otimes |0\rangle\langle 0|$,
where $\rho(0)$ represents the initial reduced density matrix of the atom and $|0\rangle$ is the vacuum state of the field. In the laboratory frame, the time evolution of the total density matrix is described by the Liouville--von Neumann equation
\begin{equation}
\frac{d \rho_{\rm tot} (t)}{d t}=-\frac{i}{\hbar}[H,\rho_{\rm tot} (t)]\;.
\end{equation}
In the weak-coupling regime between the atom and the field, the reduced dynamics of the atomic density matrix $\rho(t)=\mathrm{Tr}_F(\rho_{\rm tot})$ is governed by the Gorini-Kossakowski-Lindblad-Sudarshan master equation \cite{Kossakowski, Lindblad}, given by
 \begin{equation}\label{lindblad_eqn}
 \frac{d \rho (t)}{d t} = -\frac{i}{\hbar} [H_{\text{eff}}, \rho (t)] + \mathcal{L}[\rho (t)]\;,
 \end{equation}
with $H_{\rm eff}$ denoting the effective atomic Hamiltonian incorporating the Lamb shift and $\mathcal{L}[\rho]$ accounting for the dissipation and decoherence arising from the atom-field interaction. Since these dissipative effects are not the focus of the present work, the explicit form of $\mathcal{L}[\rho]$ will not be presented.

The Lamb shift modifies the bare atomic transition frequency $\omega$, leading to a renormalized frequency $\widetilde{\omega}$. Consequently, the effective Hamiltonian can be written as
\begin{equation}\label{HeffK}
  H_{\rm eff}=\frac{\hbar}{2} \widetilde{\omega} \sigma_3=\frac{\hbar}{2}\{\omega+\frac{i}{2}[\mathcal{K} (-\omega)-\mathcal{K} (\omega)]\} \sigma_3\;,
\end{equation}
where
\begin{equation}
\mathcal{K} (\lambda)=\sum_{\alpha ,\beta =\rho ,\phi ,z} \frac{{d}_{\alpha}{d}^*_{\beta}}{i\pi\hbar^2 }  \rm{P} \int_{-\infty}^{\infty} d \nu   \frac{\mathcal{G}_{\alpha\beta}(\nu)}{\nu-\lambda}\;,
\end{equation}
with $\alpha,\beta=\rho,\phi,z$ and ``P'' standing for the Cauchy principal value. The quantity $d_\alpha = \langle e | {d}_\alpha | g \rangle$ represents the transition matrix element of the dipole operator ${d}_\alpha$,  where $|g\rangle$ and $|e\rangle$ correspond to the ground and excited states of the atom, respectively. The function $\mathcal{G}_{\alpha\beta}(\omega)$ denotes the Fourier transform of the field correlation function and is given by
\begin{equation}
\mathcal{G}_{\alpha\beta} (\lambda)=\int_{-\infty}^{\infty} d t_{-}   e ^{i \lambda t_{-}}G_{\alpha\beta}(t_{-})\;,
\end{equation}
where $t_- = t - t'$ denotes the coordinate time interval.
In what follows, the field correlation function of the electromagnetic field in the vacuum state $|0\rangle$ is expressed as
\begin{equation}\label{Gt}
  G_{\alpha\beta}(t _{-})=\langle 0|\mathcal{E}_{\alpha}\left(t,\mathbf{x}\right)\mathcal{E}_{\beta}\left(t',\mathbf{x}'\right)|0\rangle\;,
\end{equation}
with the explicit forms of $\langle 0 | \mathcal{E}_{\alpha}(t, \mathbf{x}) \mathcal{E}_{\beta}(t', \mathbf{x}') | 0 \rangle$ provided in Appendix~\ref{sec:S2}.
Owing to time-translation invariance, the correlation function depends only on the time interval $t_-$.

According to Eq.~\eqref{HeffK}, the relative Lamb shift between the ground and excited states can be expressed as
\begin{align}\label{shift}
  \Delta =&\frac{i}{2}[\mathcal{K} (-\omega)-\mathcal{K} (\omega)]\nonumber\\
  =&\sum_{\alpha ,\beta =\rho ,\phi ,z} \frac{{d}_{\alpha}{d}^*_{\beta}}{2\pi\hbar^2 }\text{P} \int_{-\infty}^{\infty}{\text{d}}\nu \;\mathcal{G}_{\alpha\beta}\left( \nu \right) \left( \frac{1}{\nu +\omega }-\frac{1}{\nu -\omega } \right) \;.
\end{align}
Hereafter, Lamb shift will refer specifically to this relative Lamb shift between the two energy levels.

\section{Lamb shift of inertial atoms} \label{sec3}
In the laboratory frame, the trajectory of an inertial atom at rest is given by $x(t)=0,~y(t)=0$, and $z(t)=0$. The two-point correlation function of the electric field can be expressed as
\begin{align}
&\left\langle 0\left|\mathcal{E}_{\alpha}(t,\mathbf{x}) \mathcal{E}_{\beta}\left(t',\mathbf{x}'\right)\right| 0\right\rangle=\frac{\hbar}{8 \pi \epsilon_{0} V}\int_{0}^{2 \pi} d\varphi \int_{0}^{\pi} \sin\theta d\theta \nonumber\\
&\times\int_{0}^{\infty} d\omega_{k} \rho(\omega_k) \frac{\omega_k}{2} \left(\delta_{\alpha\beta}-\frac{k_{\alpha}k_{\beta}}{\boldsymbol{k}^{2}}\right) e^{-i( \omega_{k} t_{-}-\boldsymbol{k} \cdot \boldsymbol{R})}\label{correlationfunctionEE}\;,
\end{align}
where
$\epsilon_0$ denotes the vacuum permittivity, $V$ is the quantization volume, $\omega_k$ is the frequency of the field mode, and $\boldsymbol{k}$ is the wave vector.
The symbol $\delta_{\alpha\beta}$ represents the Kronecker delta. The displacement vector is defined as $\boldsymbol{R}=\mathbf{x}(t)-\mathbf{x}(t^{\prime})$,
and the quantity $\rho(\omega_k)=\frac{V \omega_k^2}{\pi^2 c^3}$ is the density of states in free space.

For an inertial atom at rest, $\boldsymbol{R}=0$, and the Fourier transform of the correlation function can be expressed as
\begin{align}
\mathcal{G}_{\alpha\beta}(\nu)=\frac{\hbar \delta_{\alpha\beta} }{3 \epsilon _0 c^3 \pi}\nu^3\;.
\end{align}
Substituting this result into Eq.~\eqref{shift}, the Lamb shift of an inertial atom in free space is obtained as
\begin{align}
  \Delta_{\rm inertial} =&		\frac{d_{\rho}^{2}+d_{\phi}^{2}+  d_{z}^{2}}{6\hbar \epsilon _0 c^3 \pi^2}\text{P}\int_{0}^{\infty}{d}\nu \left( \frac{\nu^3 }{\nu+\omega_0}-\frac{\nu^3 }{\nu-\omega_0} \right)\;\nonumber\\
  =&-\frac{d_{\rho}^{2}+d_{\phi}^{2}+  d_{z}^{2}}{3\hbar \epsilon _0 c^3 \pi^2}\omega_0 \text{P}\int_{0}^{\infty}{d}\nu \frac{\nu^3 }{\nu^2-\omega_0^2}\;.\label{Lambshiftin}
\end{align}
To evaluate the frequency integral in Eq.~\eqref{Lambshiftin}, we first decompose the integrand as
\begin{align}\label{renormalize}
\int_{0}^{\infty}{d}\nu \frac{\nu^3 }{\nu^2-\omega_0^2}=\int_{0}^{\infty}{d}\nu\nu+\omega_0^2\int_{0}^{\infty}{d}\nu \frac{\nu}{\nu^2-\omega_0^2}\;.
\end{align}
The first term leads to a power-law ultraviolet divergence that is independent of the atomic transition frequency and can, therefore, be absorbed into mass renormalization \cite{Bethe47,Milonni94}. The second term gives a logarithmically divergent contribution. Introducing an ultraviolet cutoff $\Lambda=m_e c^2 / \hbar$, where $m_e$ is the electron rest mass and $\Lambda \gg \omega_0$, we obtain
\begin{align}\label{cutoff}
-\frac{d_{\rho}^{2}+d_{\phi}^{2}+  d_{z}^{2}}{3\hbar \epsilon _0 c^3 \pi^2}\omega_0^3\int_{0}^{\Lambda}{d}\nu \frac{\nu}{\nu^2-\omega_0^2}\;.
\end{align}

As a result, the inertial Lamb shift takes the standard Bethe-type logarithmic form:
\begin{align}
  \Delta_{\rm inertial}   =&-\frac{ d_{\rho}^{2}+d_{\phi}^{2}+  d_{z}^{2}}{6\hbar \epsilon _0 c^3 \pi^2}\omega_0^3 \ln  \frac{\Lambda ^2-\omega_0 ^2}{\omega_0 ^2}\;.
\end{align}
As the cutoff frequency remains substantially greater than the atomic transition frequency, this expression simplifies to
\begin{align}\label{Lambshiftin-result}
  \Delta_{\rm inertial}   =&-\frac{ d_{\rho}^{2}+d_{\phi}^{2}+  d_{z}^{2}}{3\hbar \epsilon _0 c^3 \pi^2}\omega_0^3 \ln  \frac{\Lambda}{\omega_0}\;.
\end{align}
Equation~(\ref{Lambshiftin-result}) corresponds exactly to the result originally derived by Bethe in Ref.~\cite{Bethe47}.

\vspace{0.4em}

\section{Lamb shift of centripetally accelerated atoms in  the nonrelativistic regime} \label{sec4}

We consider a two-level atom undergoing centripetal acceleration and coupled to electromagnetic vacuum fluctuations. The atom follows a circular trajectory, $x(t)=R\cos(\Omega t),~y(t)=R\sin(\Omega t)$, and $z(t)=0$, where $R$ is the radius of the circular orbit and $\Omega$ is the angular velocity.

To evaluate the Lamb shift, we employ the field correlation functions in the laboratory frame, whose explicit forms are provided in Eqs.~\eqref{EErho}--\eqref{EEz} in Appendix~\ref{sec:S2}. After performing the Fourier transform, the resulting expressions are substituted into Eq.~\eqref{shift}. In the nonrelativistic limit, where the linear velocity satisfies $v = R\Omega \ll c$, and, retaining terms up to second order in $R$, we obtain the Lamb shift of a centripetally accelerated atom in free space as observed in the laboratory frame, which is given by
\begin{align}\label{Lambshift-free-cir}
\Delta=\sum_{\alpha=\rho,\phi,z}\frac{{d}_{\alpha}^2}{6 \epsilon _0 \hbar }\text{P} \int_{-\infty}^{\infty}{\text{d}}\nu \;\mathcal{F}_{\alpha}(\nu) \left( \frac{1}{\nu +\omega }-\frac{1}{\nu -\omega } \right)\;,
\end{align}
where the explicit integral expressions for $\mathcal{F}_{\alpha}$ are given below:
\begin{widetext}
\begin{align}
&\mathcal{F}_{\rho}(\nu)= \int_{0}^{\infty}d \omega_{k}\frac{\omega^3_{k}}{c^3 \pi^2} \nn\\
&\times\Biggl(\frac{1}{2} \left[\delta \left(\nu+\Omega -\omega _k\right)+ \delta \left(\nu -\Omega -\omega _k\right)\right]+\frac{R^2 \Omega ^2 }{c^2}\delta \left(\nu-\omega _k \right)+\frac{R^2 \Omega  \omega _k}{2 c^2}[\delta \left(\nu -\Omega -\omega _k\right)-\delta \left(\nu +\Omega -\omega _k\right)]+\frac{R^2 \omega _k^2}{20 c^2}\nn\\
&\times\left\{\delta \left(\nu-\omega _k \right)-2[\delta \left(\nu +\Omega -\omega _k\right)+\delta \left(\nu -\Omega -\omega _k\right)]+\frac{3}{2}[\delta \left(\nu -2 \Omega -\omega _k\right)+\delta \left(\nu +2 \Omega -\omega _k\right)]\right\}\Biggl) +\mathcal{O}\left[\left(\frac{R\Omega}{c}\right)^4\right]\;,\\
& \mathcal{F}_{\phi}(\nu)= \int_{0}^{\infty}d \omega_{k}\frac{\omega^3_{k}}{c^3 \pi^2} \Biggl(\frac{1}{2} [\delta \left(\nu +\Omega -\omega _k\right)+ \delta \left(\nu-\Omega -\omega _k\right)]-\frac{R^2 \Omega ^2}{2 c^2}[\delta \left(\nu+\Omega -\omega _k\right)+\delta \left(\nu -\Omega -\omega _k\right)]
+\frac{R^2 \omega _k^2}{c^2}\nn\\
&\times\left\{\frac{1}{4} \delta \left(\nu-\omega _k \right)-\frac{1}{5}[\delta \left(\nu +\Omega -\omega _k\right)+\delta \left(\nu -\Omega -\omega _k\right)]+\frac{3 }{40}[\delta \left(\nu-2 \Omega -\omega _k\right)+\delta \left(\nu +2 \Omega -\omega _k\right)]\right\}\Biggl) +\mathcal{O}\left[\left(\frac{R\Omega}{c}\right)^4\right]\;,\end{align}
and
\begin{align}
&\mathcal{F}_{z}(\nu)= \int_{0}^{\infty}d \omega_{k}\frac{\omega^3_{k}}{c^3 \pi^2} \Biggl(\delta ( \nu-\omega_{k})+\frac{R^2 \Omega ^2 }{2 c^2}\left[\delta \left(\nu +\Omega -\omega _k\right)+\delta \left(\nu -\Omega -\omega _k\right)\right]+\frac{R^2 \Omega  \omega _k}{2 c^2}\left[\delta \left(\nu -\Omega -\omega _k\right)\right.\nn\\
    &\left.-\delta \left(\nu +\Omega -\omega _k\right)\right]-\frac{2 R^2 \omega _k^2}{5 c^2}\left\{\delta \left(\nu -\omega _k\right)-\frac{1}{2}\left[\delta \left(\nu +\Omega -\omega _k\right)+\delta \left(\nu -\Omega -\omega _k\right)\right]\right\}\Biggl) +\mathcal{O}\left[\left(\frac{R\Omega}{c}\right)^4\right]\;.
\end{align}
Terms denoted by $\mathcal{O}[x^n]$ represent contributions of the order of $x^n$ or higher, which are negligibly small and, thus, neglected in the analysis. After performing the integration over $\omega_k$, the functions $\mathcal{F}_\alpha(\nu)$ can be written explicitly as
\begin{align}
&\mathcal{F}_{\rho}(\nu)= \nn\\
&\frac{1}{c^3 \pi^2}\Biggl(\frac{1}{2} \left[ \left(\nu+\Omega \right)^3 \Theta\left(\nu+\Omega \right)+ \left(\nu-\Omega \right)^3 \Theta\left(\nu-\Omega \right)\right]+\frac{R^2 \Omega ^2 }{c^2}\nu^3 \Theta(\nu)+\frac{R^2 \Omega }{2 c^2}\left[\left(\nu -\Omega \right)^4\Theta\left(\nu -\Omega \right)
-\left(\nu +\Omega \right)^4\Theta\left(\nu +\Omega \right)\right]\nn\\
&+\frac{R^2}{20 c^2}\left\{\nu^5\Theta(\nu)-2\left[\left(\nu +\Omega \right)^5\Theta\left(\nu +\Omega \right)+\left(\nu -\Omega \right)^5\Theta\left(\nu -\Omega \right)\right]+\frac{3}{2}\left[\left(\nu -2\Omega \right)^5\Theta\left(\nu -2\Omega \right)+\left(\nu +2\Omega \right)^5\Theta\left(\nu +2\Omega \right)\right]\right\}\Biggl)  \nn\\
&+\mathcal{O}\left[\left(\frac{R\Omega}{c}\right)^4\right]\;,\\
& \mathcal{F}_{\phi}(\nu)= \frac{1}{c^3 \pi^2}\Biggl(\frac{1}{2} \left[ \left(\nu+\Omega \right)^3 \Theta\left(\nu+\Omega \right)+ \left(\nu-\Omega \right)^3 \Theta\left(\nu-\Omega \right)\right]-\frac{R^2 \Omega ^2}{2 c^2}\left[\left(\nu+\Omega \right)^3\Theta\left(\nu+\Omega \right)+\left(\nu-\Omega \right)^3\Theta\left(\nu-\Omega \right)\right]+\frac{R^2 }{c^2}\nn\\
&\times\left\{\frac{1}{4}\nu^5\Theta(\nu)-\frac{1}{5}\left[ \left(\nu +\Omega \right)^5\Theta\left(\nu +\Omega \right)+ \left(\nu -\Omega \right)^5\Theta\left(\nu -\Omega \right)\right]+\frac{3 }{40}\left[\left(\nu-2 \Omega \right)^5\Theta\left(\nu-2 \Omega \right)+\left(\nu+2 \Omega \right)^5\Theta\left(\nu+2 \Omega \right)\right]\right\}\Biggl)  \nn\\
&+\mathcal{O}\left[\left(\frac{R\Omega}{c}\right)^4\right]\;,
\end{align}
and
\begin{align}
&\mathcal{F}_{z}(\nu)= \frac{1}{c^3 \pi^2}\Biggl(\nu^3\Theta(\nu)+\frac{R^2 \Omega ^2 }{2 c^2}\left[ \left(\nu +\Omega\right)^3\Theta\left(\nu +\Omega\right)+\left(\nu -\Omega \right)^3\Theta\left(\nu -\Omega \right)\right]+\frac{R^2 \Omega  }{2 c^2}\left[\left(\nu -\Omega \right)^4\Theta\left(\nu -\Omega \right)\right.\nn\\
&\left.-\left(\nu +\Omega \right)^4\Theta\left(\nu +\Omega \right)\right]-\frac{2 R^2 }{5 c^2}\left\{\nu^5\Theta(\nu)-\frac{1}{2}\left[ \left(\nu +\Omega \right)^5\Theta\left(\nu +\Omega \right)+ \left(\nu -\Omega \right)^5\Theta\left(\nu -\Omega \right)\right]\right\}\Biggl) +\mathcal{O}\left[\left(\frac{R\Omega}{c}\right)^4\right]\;,
\end{align}
\end{widetext}
where the Heaviside step function $\Theta(x)$ equals zero for $x<0$ and unity for $x>0$.

Following the same procedure used for the Lamb shift of an inertial atom in free space [see Eqs.~\eqref{renormalize} and \eqref{cutoff}], we renormalize Eq. \eqref{Lambshift-free-cir}, retaining only the terms that contribute to the Lamb shift and introducing an ultraviolet cutoff $\Lambda$ with $\Lambda\gg \omega$ and $\Lambda\gg \Omega$ as the upper limit of integration. This procedure yields the Lamb shift of a centripetally accelerated atom in free space, as observed in the laboratory frame in the nonrelativistic limit:
\begin{widetext}
\begin{align}\label{circular-Lambshift}
 \Delta=\Delta_{\rm inertial}+\Delta_{\rm circular}^{(0)}+\Delta_{\rm circular}^{(2)}+\mathcal{O}\left[\left(\frac{R\Omega}{c}\right)^4\right]\;,
\end{align}
where
\begin{align}\label{circular-Lambshift-0}
\Delta_{\rm circular}^{(0)}=&-\frac{d_{\rho}^2+d_{\phi}^2}{6 \pi ^2 c^3 \epsilon _0 \hbar }\left[\frac{(\omega_0 -\Omega )^3 }{2} \ln\frac{\Lambda ^2-\omega_0 ^2}{(\omega_0 -\Omega )^2}+\frac{(\omega_0 +\Omega )^3}{2} \ln\frac{\Lambda ^2-\omega_0 ^2}{(\omega_0 +\Omega )^2}+2 \omega_0  \Omega ^2 \left(1-3 \ln\frac{\Lambda }{\Omega }\right)-\omega_0 ^3 \ln \frac{\Lambda ^2-\omega_0 ^2}{\omega _0^2}\right]\;
\end{align}
is the zeroth-order rotational correction to the Lamb shift and the second-order rotational contribution reads
\begin{align}\label{circular-Lambshift-2}
\Delta_{\rm circular}^{(2)}=&\frac{R^2}{6 \pi ^2 c^5 \epsilon _0 \hbar }\left(-\frac{1}{5} \omega_0  \Omega ^2 \left[4 \left(2 \omega_0 ^2+\Omega ^2\right) \left(d_{\rho }^2+d_{\phi }^2+d_z^2\right)+3 \Omega ^2 (1+40 \ln 2) \left(d_{\rho }^2+d_{\phi }^2\right)+\left(\omega_0 ^2-2 \Omega ^2\right)d_{\phi }^2 \right]\right.\nonumber\\
&\left.+ \omega_0  \Omega ^4 \left(11 d_{\rho }^2+11d_{\phi }^2+2d_z^2\right)\ln \frac{\Lambda }{\Omega }-\frac{1}{20}\left[\left(\omega_0 ^5+20 \omega_0 ^3 \Omega ^2\right)d_{\rho }^2 +5 \omega_0 ^5 d_{\phi }^2-2 \left(4 \omega_0^5+15 \omega_0^3 \Omega ^2\right) d_z^2\right] \ln \frac{\Lambda ^2-\omega_0 ^2}{\omega_0 ^2} \right.\nonumber\\
&\left.+\sum_{p=\pm 1}\left\{\frac{(\omega_0 +p\Omega )^3}{10} \left[ (\omega_0 ^2+7p \omega_0  \Omega +6 \Omega ^2 )d_{\rho }^2+ \left(2 \omega_0 ^2+4 p\omega _0 \Omega +7 \Omega ^2\right)d_{\phi }^2-\left(2 \omega_0 ^2-p\omega_0  \Omega +2 \Omega ^2\right)d_z^2 \right]\right.\right.\nonumber\\
&\left.\left.+\frac{3\omega_0 \Omega ^2}{4}  (\omega_0+p\Omega )^2(d_{\rho }^2+d_{\phi }^2)\right\}\ln\frac{\Lambda ^2-\omega_0 ^2}{(\omega_0 +p\Omega )^2}-\frac{3}{40} \sum_{p=\pm 1}(\omega_0 +2 p\Omega )^5 \left(d_{\rho }^2+d_{\phi }^2\right) \ln\frac{\Lambda ^2-\omega_0 ^2}{(\omega_0 +2p \Omega )^2}
\right)\;.
\end{align}
\end{widetext}
From Eqs.~\eqref{circular-Lambshift-0} and \eqref{circular-Lambshift-2}, it is evident that the zeroth-order correction arises entirely from transverse ($\rho$ and $\phi$) polarizations, whereas the second-order term also receives contributions from the axial ($z$) polarization.

These results show that the polarization components contribute to the Lamb shift in an intrinsically anisotropic way. This contrasts with the case of uniformly linearly accelerated atoms, for which all polarization directions contribute equally~\cite{Passante98}. In the following, we therefore analyze separately the contributions to the Lamb shift from axial polarization ($z$ direction) and transverse polarization ($\rho$ and $\phi$ directions).
All subsequent discussions are based on the nonrelativistic result obtained under the condition $R\Omega/c\ll1$, with the small-radius expansion retained through second order in $R$, as given in Eqs.~\eqref{circular-Lambshift}--\eqref{circular-Lambshift-2}. The explicit axial and transverse contributions, together with the details of the asymptotic expansions employed in the small- and large-angular-velocity regimes, are provided in Appendix~\ref{sec:S3}.

\subsection{Axial-polarization contribution to the Lamb shift}\label{sec4:A}
We first consider the contribution to the Lamb shift  from the atomic polarization along the rotation axis, denoted by $\Delta^{\parallel}$. Its explicit expression is given in Eq.~\eqref{circular-Lambshift-parallel} in Appendix~\ref{sec:S3}. 
Under the small-angular-velocity approximation $(\Omega\ll\omega_0\ll\Lambda)$, the axial contribution to the Lamb shift of a centripetally accelerated atom in free space can be approximated by
\begin{align}\label{z-small-velocity-Lambshift}
  \Delta^{\parallel} \approx -\frac{\omega_0 ^3 d_z^2}{3 \pi ^2 c^3 \epsilon _0 \hbar }\ln \frac{\Lambda }{\omega_0 }\left(1-\frac{R^2 \Omega ^2}{2 c^2}\right)\;.
\end{align}
The leading term corresponds to the standard Lamb shift of an inertial atom in free space, while the second term represents the rotational correction due to centripetal acceleration. This correction is $R^2\Omega^2/2c^2$ times the inertial Lamb shift, but its sign is opposite. Consequently, rotation makes the negative Lamb shift slightly less negative and, therefore, produces a small increase in the energy-level spacing.

In the opposite limit of large angular velocity  $(\omega_0\ll\Omega\ll\Lambda)$, the axial contribution becomes
\begin{align}\label{z-big-velocity-Lambshift}
  \Delta^{\parallel}  &\approx-\frac{\omega_0 ^3 d_z^2}{3 \pi ^2 c^3 \epsilon _0 \hbar }\left[\ln\frac{\Lambda }{\omega_0 }+\frac{R^2 \Omega ^2 }{c^2}\right.\nonumber\\
  &\;\;\;\;\;\;\;\;\;\;\;\;\;\;\;\;\;\;\;\;\; \left.\times\left(\ln\frac{\omega_0}{\Omega }-\frac{1}{2} \ln \frac{\Lambda }{\omega_0}+\frac{1}{6}\right)\right]\;.
\end{align}
Again, the first term in Eq.~(\ref{z-big-velocity-Lambshift}) remains identical to the inertial Lamb shift.
The second rotation-induced  term scales as $\frac{R^2 \Omega ^2 }{c^2}\left(\ln\frac{\omega_0}{\Omega }-\frac{1}{2} \ln \frac{\Lambda }{\omega_0}+\frac{1}{6}\right)$. Although it grows quadratically with the angular velocity $\Omega$, it is suppressed by the prefactor $(R^2 \Omega^2 / c^2)$ as we work in the nonrelativistic regime ($R\Omega \ll c$). Consequently, the rotational contribution remains a small, subleading correction to the total Lamb shift. Moreover, since here $\omega_0\ll\Omega\ll\Lambda$, the bracket is negative, so the rotational correction slightly reduces the Lamb shift or, equivalently, slightly increases the energy-level spacing.

\subsection{Transverse-polarization contribution to the Lamb shift}\label{sec4:B}
We now turn to the contribution from transverse polarization, namely, the case in which the atom is polarizable along a direction perpendicular to the rotation axis, denoted by $\Delta^{\perp}$. The explicit expression is shown in Eq.~\eqref{circular-Lambshift-perp} in Appendix~\ref{sec:S3}. In the nonrelativistic regime and under the small-angular-velocity approximation $\Omega \ll \omega_0\ll\Lambda$, the transverse contribution to the Lamb shift of a centripetally accelerated atom in free space can be approximated as
\begin{align}\label{vertical-small-velocity-Lambshift}
\Delta^{\perp}  \approx&-\frac{\omega_0 ^3 \left(d_{\rho }^2+d_{\phi }^2\right)}{3 \pi ^2 c^3 \epsilon _0 \hbar }\left[\ln \frac{\Lambda }{\omega_0 }-\frac{3 \Omega ^2}{2 \omega_0 ^2}\left(2 \ln \frac{\omega_0 }{\Omega }+1\right)\right]\;.
\end{align}
The first leading term in Eq.~(\ref{vertical-small-velocity-Lambshift}) corresponds to the Lamb shift of an inertial atom, while the second term in the square brackets is the rotation-induced correction.  Therefore, in the small-angular-velocity limit, the Lamb shift is always dominated by the inertial contribution, irrespective of the polarization orientation.

A key difference from the axial case is that, for transverse polarization, the leading rotational correction  is independent of the orbital radius $R$; it thus enters already at zeroth order in an expansion with respect to $R$. Even so, it is still small  in the present regime, as it is suppressed by the factor $\frac{\Omega^2}{\omega_0^2}$.
Since $\Omega<\omega_0$, the bracket is positive, implying that the rotational correction slightly reduces the Lamb shift or, equivalently, slightly increases the energy-level spacing.

In contrast, in the opposite regime of  large angular velocity ($\omega_0\ll\Omega \ll \Lambda$), the transverse contribution exhibits a qualitatively different dependence on the angular velocity and takes the form
\begin{align}\label{vertical-big-velocity-Lambshift}
  \Delta^{\perp}  \approx &-\frac{\omega_0 ^3 \left(d_{\rho }^2+d_{\phi }^2\right)}{3 \pi ^2 c^3 \epsilon _0 \hbar }\left[\ln\frac{\Lambda }{\omega_0 }+\left( \ln\frac{\omega_0 }{\Omega }-\frac{11}{6}\right)\right]
  \;.
\end{align}
As before, the first term corresponds to the Lamb shift of an inertial atom, while the second term in the square brackets is the rotation-induced correction. Interestingly, this rotational correction includes a constant term that is independent of the angular velocity.
In the large-angular-velocity limit, the Lamb shift depends on $\Omega$ primarily through a logarithmic factor. An increase in $\Omega$, therefore, decreases the magnitude of the transverse contribution only logarithmically rather than via any power-law scaling.
Since $\Omega  >\omega_0$ in this regime, $\ln(\omega_0/\Omega)$ is negative, and the bracketed correction is negative overall.
Therefore, rotation again makes $\Delta^{\perp} $ less negative, corresponding to an increased energy-level spacing.

\vspace{0.8em}

Now, we estimate the relative weight of the rotation-induced correction to the Lamb shift. In practice, atoms are generally isotropically polarizable, such that $d_\rho^2=d_\phi^2=d_z^2$. Typically, the atomic energy-level spacing $\hbar \omega_0$ is of the order of 1 eV, while the ultraviolet cutoff $\hbar \Lambda$ is set by the electron rest mass, about 0.5 MeV. Assuming further that the angular velocity satisfies $\Omega \sim 10\omega_0$ and that the orbital radius is sufficiently small so that only the zeroth-order rotational correction in $R$ needs to be retained, the ratio of the rotational contribution [Eq.~\eqref{circular-Lambshift-0}] to the inertial one [Eq.~\eqref{Lambshiftin-result}] can exceed $20\%$. In this sense, the rotational correction may become comparable in magnitude to the inertial Lamb shift even when the associated centripetal acceleration is extremely small due to a sufficiently small orbital radius. This highlights the nontrivial role of rotational motion in reshaping vacuum-induced radiative energy shifts.

\section{Summary} \label{sec5}
In this work, we systematically investigate the Lamb shift of centripetally accelerated atoms in the electromagnetic vacuum of free space. We focus on the  nonrelativistic  regime of a very small orbital radius, such that the linear speed is much smaller than the speed of light, and the centripetal acceleration is also extremely small. Despite this, we find that circular motion can induce appreciable, strongly anisotropic corrections to the Lamb shift, controlled by the direction of the atomic polarizability relative to the rotation axis.

For atoms polarizable \emph{along} the rotation axis, the leading noninertial contribution enters at second order in the orbital radius. In both the low- and high-angular-velocity regimes, it produces a small increase in the energy-level spacing.

In contrast, for atoms polarizable \emph{perpendicular} to the rotation axis, the rotation-induced contribution appears already at zeroth order in the orbital radius. In the low-angular-velocity regime, it increases the energy-level spacing. In the high-angular-velocity regime, the correction separates into two distinct pieces: one that depends explicitly on the angular velocity and another that is independent of it. Remarkably, these rotational contributions can become comparable to the inertial Lamb shift, implying a substantial modification even in the limit of a vanishing orbital radius---and, hence, a vanishing proper centripetal acceleration. This highlights circular motion as a route to amplifying noninertial signatures in precision level-shift observables through polarization-selective vacuum correlations.

Our results indicate that the Lamb shift can serve as a sensitive probe of acceleration-induced modifications of the quantum vacuum, and may be accessible to experimental verification through high-precision spectroscopy.

\begin{acknowledgments}
This work was supported in part by the NSFC under Grants No. 12075084 and No. 12575051 and the innovative research group of Hunan Province under Grant No. 2024JJ1006.
\end{acknowledgments}

\section*{Data Availability}

There are no publicly available research data or software supporting
this manuscript. Requests for further information or data should be
sent to the authors.

\renewcommand{\appendixname}{APPENDIX}
\appendix

\section{THE INTERACTION HAMILTONIAN IN THE LABORATORY FRAME}\label{sec:S1}
The dipole interaction between an atom and the electromagnetic field in the proper frame of the atom can be expressed in a coordinate-invariant form as~\cite{Takagi86}
\begin{equation}\label{interaction1}
H_{I}= - {d}^{\prime \mu} F_{\mu\nu} u^{\nu}\;,
\end{equation}
where ${d}^{\prime\mu}$ represents the four-electric dipole moment operator of the atom, which is $(0,{d}_{\rho},{d}_{\phi},{d}_z)$ in the proper frame of the atom. $F_{\mu\nu}$ is the electromagnetic tensor, and
$u^{\nu}$ is the four-velocity of the atom.

To express the Hamiltonian in the laboratory frame, one introduces the Lorentz factor $\gamma=(1- \beta^2)^{-\frac{1}{2}}$, yielding
\begin{equation}\label{interaction1}
H_{I}= - \frac{1}{\gamma}{d}^{\mu} F_{\mu\nu} u^{\nu}\;.
\end{equation}
Here, $\beta=R \Omega/c$, where $R$ and $\Omega$ denote the radius and rotational angular velocity of the atom, respectively, and $c$ is the speed of light.

To express the dipole moment operator in the laboratory frame ${d}^{\mu}$
in terms of its components in the proper frame ${d}^{\prime \mu}=(0,{d}_{\rho},{d}_{\phi},{d}_z)$, a combination of rotational transformation and Lorentz transformation is necessary, i.e.,
\begin{equation}\label{trans}
{d}^{\mu}
=\Lambda_{j}^{\mu} S_{i}^{j} {d}^{\prime i},
\end{equation}
where
\begin{eqnarray}\label{Ax}
\Lambda_{j}^\mu=\left(
  \begin{array}{cccc}
    \gamma                   & -\gamma \beta   n_x  &         -\gamma \beta n_y          &0 \\
    -\gamma \beta n_x  & 1+(\gamma-1)n_x^2         &         (\gamma-1)n_x n_y                & 0\\
   -\gamma \beta n_y   &    (\gamma-1)n_y n_x      &         1+(\gamma-1)n_y^2                & 0\\
      0                      &          0                &         0                                &  1\\
  \end{array}
\right)~~
\end{eqnarray}
is the Lorentz transformation matrix, and
\begin{eqnarray}
S_i^j=
\left(
  \begin{array}{cccc}
     n_y &  n_x        &0\\
   -n_x   &  n_y        & 0\\
      0                     &  0                              & 1
  \end{array}
\right)\;
\end{eqnarray}
is the rotational transformation matrix.
Here, $n_x=v_x/v=-\sin(\Omega t)$ and $n_y=v_y/v=\cos(\Omega t)$ are the components of the unit vector along the direction of the velocity, with $t$ being the coordinate time.
In the laboratory frame, the electromagnetic tensor $F_{\mu\nu}$ takes the form
\begin{eqnarray}\label{Fuv}
F_{\mu\nu}=
\left(
  \begin{array}{cccc}
     0         & -E_x /c       & -E_y /c     & -E_z /c   \\
E_x /c         & 0           & B_z      & -B_y\\
E_y /c         & -B_z        & 0        & B_x\\
E_z /c         & B_y         & -B_x     & 0
  \end{array}
\right)\;,
\end{eqnarray}
where $E_i$ and $B_i$ $(i=x,y,z)$ are the components of the electric and magnetic fields in the laboratory frame, respectively.
The four-velocity of the atom in the laboratory frame is given by
\begin{align}\label{uv}
    u^{\nu}=c \gamma (1, \beta n_x,\beta n_y,0).
\end{align}
Taking Eqs.~\eqref{trans}, \eqref{Fuv} and \eqref{uv} into Eq.~\eqref{interaction1}, the interaction Hamiltonian in the laboratory frame can be written as,
\begin{align}\label{HI-lab}
H_{I}=&-{d}_{\rho}\left[\Omega R B_{z} +\cos(\Omega t) E_{x} +\sin(\Omega t) E_y   \right]\nonumber\\
&-\frac{{d}_{\phi}}{\gamma} \left[ \cos(\Omega t)  E_{y} - \sin(\Omega t) E_{x}\right]\nonumber\\
&-{d}_{z} \left[E_{z}   -\Omega R \sin(\Omega t)B_{y}  -\Omega R \cos(\Omega t)B_{x}\right]\;.
\end{align}
\vspace{-8pt}

\section{TWO-POINT FUNCTIONS OF THE FLUCTUATING ELECTROMAGNETIC FIELDS IN FREE SPACE}\label{sec:S2}

The Lamb shift is dependent on the two-point correlation functions of the electromagnetic fields $G_{\alpha\beta}(t_{-})=\langle 0|\mathcal{E}_{\alpha}\left(t,\mathbf{x}\right)\mathcal{E}_{\beta}\left(t',\mathbf{x}'\right)|0\rangle$. For simplicity, assuming the dipole matrix elements ${d}_{\alpha}$ are real, the cross terms in Eq.~\eqref{Gt} vanish. The diagonal components of the two-point functions are then given by~\cite{yu26}
\begin{widetext}
\begin{eqnarray}\label{EErho}
\langle 0|\mathcal{E}_{\rho}\left(t,\mathbf{x}\right)\mathcal{E}_{\rho}\left(t',\mathbf{x}'\right) |0\rangle
&=& \Omega^2 R^2\langle 0|B_{z}\left(t,\mathbf{x}\right)B_{z}\left(t',\mathbf{x}'\right) |0\rangle+ \Omega R \cos(\Omega t')\langle 0|B_{z}\left(t,\mathbf{x}\right)E_{x}\left(t',\mathbf{x}'\right)|0\rangle\nonumber\\
&&+ \Omega R \sin(\Omega t')\langle 0|B_{z}\left(t,\mathbf{x}\right)E_{y}\left(t',\mathbf{x}'\right) |0\rangle+ \Omega R \cos(\Omega t)\langle 0|E_{x}\left(t,\mathbf{x}\right)B_{z}\left(t',\mathbf{x}'\right) |0\rangle\nonumber\\
&&+ \cos(\Omega t)\cos(\Omega t')\langle 0|E_{x}\left(t,\mathbf{x}\right)E_{x}\left(t',\mathbf{x}'\right)|0\rangle
+ \cos(\Omega t)\sin(\Omega t')\langle 0|E_{x}\left(t,\mathbf{x}\right)E_{y}\left(t',\mathbf{x}'\right) |0\rangle\nonumber\\
&&+ \Omega R \sin(\Omega t)\langle 0|E_{y}\left(t,\mathbf{x}\right)B_{z}\left(t',\mathbf{x}'\right) |0\rangle
+ \sin(\Omega t)\cos(\Omega t')\langle 0|E_{y}\left(t,\mathbf{x}\right)E_{x}\left(t',\mathbf{x}'\right)|0\rangle\nonumber\\
&&+ \sin(\Omega t)\sin(\Omega t')\langle 0|E_{y}\left(t,\mathbf{x}\right)E_{y}\left(t',\mathbf{x}'\right) |0\rangle\;,
\end{eqnarray}
\begin{align}\label{EEphi}
\langle 0|\mathcal{E}_{\phi}\left(t,\mathbf{x}\right)\mathcal{E}_{\phi}\left(t',\mathbf{x}'\right)|0\rangle
=&\gamma^{-2}\left[\cos(\Omega t)\cos(\Omega t')\langle 0|E_{y}\left(t,\mathbf{x}\right)E_{y}\left(t',\mathbf{x}'\right) |0\rangle
-\cos(\Omega t)\sin(\Omega t')\langle 0|E_{y}\left(t,\mathbf{x}\right)E_{x}\left(t',\mathbf{x}'\right) |0\rangle\right.\nonumber\\
&\left.-\sin(\Omega t)\cos(\Omega t')\langle 0|E_{x}\left(t,\mathbf{x}\right)E_{y}\left(t',\mathbf{x}'\right) |0\rangle
+\sin(\Omega t)\sin(\Omega t')\langle 0|E_{x}\left(t,\mathbf{x}\right)E_{x}\left(t',\mathbf{x}'\right) |0\rangle\right]\;,
\end{align}
\begin{eqnarray}\label{EEz}
\langle 0|\mathcal{E}_{z}\left(t,\mathbf{x}\right)\mathcal{E}_{z}\left(t',\mathbf{x}'\right)|0\rangle
&=& \langle 0|E_{z}\left(t,\mathbf{x}\right)E_{z}\left(t',\mathbf{x}'\right) |0\rangle
- \Omega R\sin(\Omega t')\langle 0|E_{z}\left(t,\mathbf{x}\right)B_{y}\left(t',\mathbf{x}'\right) |0\rangle\nonumber\\
&&- \Omega R\cos(\Omega t')\langle 0|E_{z}\left(t,\mathbf{x}\right)B_{x}\left(t',\mathbf{x}'\right) |0\rangle
- \Omega R\sin(\Omega t)\langle 0|B_{y}\left(t,\mathbf{x}\right)E_{z}\left(t',\mathbf{x}'\right) |0\rangle\nonumber\\
&&+\Omega^2 R^2 \sin(\Omega t)\sin(\Omega t')\langle 0|B_{y}\left(t,\mathbf{x}\right)B_{y}\left(t',\mathbf{x}'\right) |0\rangle\nonumber\\
&&+\Omega^2 R^2 \sin(\Omega t)\cos(\Omega t')\langle 0|B_{y}\left(t,\mathbf{x}\right)B_{x}\left(t',\mathbf{x}'\right) |0\rangle\nonumber\\
&&- \Omega R\cos(\Omega t)\langle 0|B_{x}\left(t,\mathbf{x}\right)E_{z}\left(t',\mathbf{x}'\right) |0\rangle
+\Omega^2 R^2 \cos(\Omega t)\sin(\Omega t')\langle 0|B_{x}\left(t,\mathbf{x}\right)B_{y}\left(t',\mathbf{x}'\right) |0\rangle\nonumber\\
&&+\Omega^2 R^2 \cos(\Omega t)\cos(\Omega t')\langle 0|B_{x}\left(t,\mathbf{x}\right)B_{x}\left(t',\mathbf{x}'\right) |0\rangle\;,
\end{eqnarray}
where
\begin{eqnarray}
&&\left\langle 0\left|E_{l}(t,\mathbf{x}) E_{p}\left(t',\mathbf{x}'\right)\right| 0\right\rangle=\frac{\hbar}{8 \pi \epsilon_{0} V}\int_{0}^{2 \pi} d\varphi \int_{0}^{\pi} \sin\theta d\theta \int_{0}^{\infty} d\omega_{k} \rho(\omega_k) \frac{\omega_k}{2} \left(\delta_{lp}-\frac{k_{l}k_{p}}{\boldsymbol{k}^{2}}\right) e^{-i( \omega_{k} t_{-}-\boldsymbol{k} \cdot \boldsymbol{R})}\label{correlationfunctionEE}\;,\\
&&\left\langle 0\left|B_{l}(t,\mathbf{x}) B_{p}\left(t',\mathbf{x}'\right)\right| 0\right\rangle=\frac{\hbar}{8 \pi\epsilon_{0} V}\int_{0}^{2 \pi} d\varphi \int_{0}^{\pi} \sin\theta d\theta \int_{0}^{\infty} d\omega_{k} \rho(\omega_k) \frac{\omega_k}{2c^2} \left(\delta_{lp}-\frac{k_{l}k_{p}}{\boldsymbol{k}^{2}}\right)  e^{-i( \omega_{k} t_{-}-\boldsymbol{k} \cdot\boldsymbol{R})}\label{correlationfunctionBB}\;,\\
&&\left\langle 0\left|E_{l}(t,\mathbf{x}) B_{p}\left(t',\mathbf{x}'\right)\right| 0\right\rangle=\frac{\hbar}{8 \pi \epsilon_{0} V}\int_{0}^{2 \pi} d\varphi \int_{0}^{\pi} \sin\theta d\theta\int_{0}^{\infty} d\omega_{k} \rho(\omega_k) \frac{\omega_k}{2c} \epsilon_{lpq} \frac{k_{q}}{\lvert \boldsymbol{k}\rvert} e^{-i( \omega_{k} t_{-}-\boldsymbol{k} \cdot\boldsymbol{R})}\label{correlationfunctionEB}\;,\\
&&\left\langle 0\left|B_{l}(t,\mathbf{x}) E_{p}\left(t',\mathbf{x}'\right)\right| 0\right\rangle=\frac{\hbar}{8 \pi \epsilon_{0} V}\int_{0}^{2 \pi} d\varphi \int_{0}^{\pi} \sin\theta d\theta\int_{0}^{\infty} d\omega_{k} \rho(\omega_k) \frac{\omega_k}{2c} \left(-\epsilon_{lpq} \frac{k_{q}}{\lvert \boldsymbol{k}\rvert}\right)  e^{-i( \omega_{k} t_{-}-\boldsymbol{k} \cdot\boldsymbol{R})}\label{correlationfunctionBE}\;.
\end{eqnarray}
\end{widetext}
In the formulas shown above, $l,p,q=x,y,z$, $\hbar$ is the reduced Planck constant, $\epsilon_0$ is the vacuum permittivity, $V$ is the quantization volume, $\omega_k$ is the frequency of the field mode, $\boldsymbol{k}$ is the wave vector, $ \epsilon_{lpq}$ is the Levi-Civita symbol, $\boldsymbol{R}=\mathbf{x}(t)-\mathbf{x}(t^{\prime})$ is a time-dependent displacement vector as the atom undergoes rotation,
$t_{-}=t_{}-t^{\prime}$, and $\rho(\omega_k)=\frac{V \omega_k^2}{\pi^2 c^3}$ is the density of states in free space.
Although Eqs.~\eqref{EErho}--\eqref{EEz} are written as functions of $t$ and $t'$, the resulting correlation functions depend only on the time difference $t_-$, but not on $t_+=t+t'$, after the integrations are carried out, reflecting the stationarity of the circular trajectory. Specifically, the apparent $t_+$ dependence can be absorbed into a rotation of the transverse momentum variables:
\begin{eqnarray}
    k_x'&=&k_x\cos\frac{\Omega t_+}{2}+k_y\sin\frac{\Omega t_+}{2}\;,\\
    k_y'&=&-k_x\sin\frac{\Omega t_+}{2}+k_y\cos\frac{\Omega t_+}{2}\;,\\
    k_z'&=&k_z\;.
\end{eqnarray}
Equivalently, in spherical momentum coordinates, this transformation corresponds to a shift of the azimuthal mode variable:
\begin{equation}
\varphi'=\varphi-\frac{\Omega t_+}{2}\;.
\end{equation}
Since the angular integration covers a complete $2\pi$ period, the resulting integral is independent of this shift and, hence, of $t_+$. Consequently, after integration over all field modes, the Wightman functions depend only on  $t_-$.

\section{ASYMPTOTIC EXPANSIONS OF THE AXIAL AND TRANSVERSE LAMB SHIFTS}\label{sec:S3}

We start from the nonrelativistic results given in Eqs.~\eqref{circular-Lambshift}--\eqref{circular-Lambshift-2}. The Lamb shift depends anisotropically on the atomic polarization and can be decomposed as
\begin{align}
\Delta=\Delta^{\parallel}+\Delta^{\perp}\;,
\end{align}
where $\Delta^{\parallel}$ denotes the contribution from polarization along the rotation axis while $\Delta^{\perp}$ denotes that from polarization in the plane perpendicular to the rotation axis. Notably, the axial-polarization contribution contains no rotation-induced term at zeroth order in the orbital radius. In this appendix, we derive the asymptotic expressions presented in the main text for the small- and large-angular-velocity regimes, namely, Eqs.~\eqref{z-small-velocity-Lambshift}--\eqref{vertical-big-velocity-Lambshift}.

\subsection{Axial-polarization contribution}

For polarization along the rotation axis, the zeroth-order rotation-induced contribution vanishes. Consequently, the leading noninertial correction appears at second order in the orbital radius. Extracting the terms proportional to $d_z^2$ from Eqs.~\eqref{circular-Lambshift}--\eqref{circular-Lambshift-2}, we obtain

\begin{widetext}
\begin{align}\label{circular-Lambshift-parallel}
 \Delta^{\parallel}=&-\frac{ d_{z}^{2} \omega_0^3 }{3\hbar \epsilon _0 \pi^2 c^3 }\ln  \frac{\Lambda ^2-\omega_0 ^2}{\omega_0 ^2}+\frac{R^2d_z^2}{6 \hbar\epsilon _0 \pi ^2 c^5  }\left[-\frac{4}{5} \omega_0  \Omega ^2  \left(2 \omega_0 ^2+\Omega ^2\right)+2 \omega_0  \Omega ^4\ln \frac{\Lambda }{\Omega }+\frac{1}{10}\left(4 \omega_0^5+15 \omega_0^3 \Omega ^2\right)  \ln \frac{\Lambda ^2-\omega_0 ^2}{\omega_0 ^2} \right.\nonumber\\
&\left.-\sum_{p=\pm 1}\frac{(\omega_0 +p\Omega )^3}{10} \left(2 \omega_0 ^2-p\omega_0  \Omega +2 \Omega ^2\right)\ln\frac{\Lambda ^2-\omega_0 ^2}{(\omega_0 +p\Omega )^2}
\right]+\mathcal{O}\left[\left(\frac{R\Omega}{c}\right)^4\right]\;.
\end{align}

To obtain transparent analytical expressions, we perform asymptotic expansions of Eq.~\eqref{circular-Lambshift-parallel} in terms of the dimensionless parameters $\Omega/\Lambda$ and $\omega_0/\Lambda$. In the small-angular-velocity regime, $\Omega\ll\omega_0\ll\Lambda$, corresponding to $\Omega/\Lambda\ll\omega_0/\Lambda\ll1$, the expressions are expanded successively about $\Omega/\Lambda\sim 0$ up to second order and subsequently around $\omega_0/\Lambda\sim0$ up to third order. The resulting expression is
\begin{align}\label{}
  \Delta^{\parallel} \approx -\frac{\omega_0 ^3 d_z^2}{3 \pi ^2 c^3 \epsilon _0 \hbar }\ln \frac{\Lambda }{\omega_0 }\left(1-\frac{R^2 \Omega ^2}{2 c^2}\right)\;,
\end{align}
which is Eq.~\eqref{z-small-velocity-Lambshift} in the main text.

In the large-angular-velocity regime, $\omega_0\ll\Omega\ll\Lambda$, corresponding to $\omega_0/\Lambda\ll\Omega/\Lambda\ll1$, the expansions are carried out in the reverse order, first with respect to $\omega_0/\Lambda$ and then with respect to $\Omega/\Lambda$.  This yields
\begin{align}\label{}
  \Delta^{\parallel}  &\approx-\frac{\omega_0 ^3 d_z^2}{3 \pi ^2 c^3 \epsilon _0 \hbar }\left[\ln\frac{\Lambda }{\omega_0 }+\frac{R^2 \Omega ^2 }{c^2}
  \left(\ln\frac{\omega_0}{\Omega }-\frac{1}{2} \ln \frac{\Lambda }{\omega_0}+\frac{1}{6}\right)\right]\;,
\end{align}
which is Eq.~\eqref{z-big-velocity-Lambshift} in the main text. The corresponding physical behavior is discussed in detail in  Sec.~\ref{sec4:A}.

\subsection{Transverse-polarization contribution}

We next consider polarization perpendicular to the rotation axis. In contrast to the axial case, the leading rotation-induced contribution is already present at zeroth order in the orbital radius. Retaining this leading small-radius contribution, we obtain
\begin{align}\label{circular-Lambshift-perp}
 \Delta^{\perp}=&-\frac{ d_{\rho}^{2}+d_{\phi}^{2}}{6\hbar \epsilon _0 \pi^2 c^3 }\left[\frac{(\omega_0 -\Omega )^3 }{2} \ln\frac{\Lambda ^2-\omega_0 ^2}{(\omega_0 -\Omega )^2}+\frac{(\omega_0 +\Omega )^3}{2} \ln\frac{\Lambda ^2-\omega_0 ^2}{(\omega_0 +\Omega )^2}+2 \omega_0  \Omega ^2 \left(1-3 \ln\frac{\Lambda }{\Omega }\right)\right]
+\mathcal{O}\left[\left(\frac{R\Omega}{c}\right)^2\right]\;.
\end{align}

Following the asymptotic analysis presented above for the axial-polarization contribution, we simplify Eq.~\eqref{circular-Lambshift-perp} in the two relevant angular-velocity regimes. In the small-angular-velocity regime, $\Omega\ll\omega_0\ll\Lambda$, we obtain
\begin{align}
\Delta^{\perp}  \approx&-\frac{\omega_0 ^3 \left(d_{\rho }^2+d_{\phi }^2\right)}{3 \pi ^2 c^3 \epsilon _0 \hbar }\left[\ln \frac{\Lambda }{\omega_0 }-\frac{3 \Omega ^2}{2 \omega_0 ^2}\left(2 \ln \frac{\omega_0 }{\Omega }+1\right)\right]\;,
\end{align}
which agrees with Eq.~\eqref{vertical-small-velocity-Lambshift} in the main text.

In the large-angular-velocity regime, $\omega_0\ll\Omega\ll\Lambda$, the corresponding expression becomes
\begin{align}
  \Delta^{\perp}  \approx &-\frac{\omega_0 ^3 \left(d_{\rho }^2+d_{\phi }^2\right)}{3 \pi ^2 c^3 \epsilon _0 \hbar }\left[\ln\frac{\Lambda }{\omega_0 }+\left( \ln\frac{\omega_0 }{\Omega }-\frac{11}{6}\right)\right]
  \;,
\end{align}
which yields Eq.~\eqref{vertical-big-velocity-Lambshift} in the main text. The physical behavior implied by these expressions is discussed in detail in Sec.~\ref{sec4:B}.

\end{widetext}

\end{document}